\documentclass[twoside,twocolumn,9pt]{article}
\usepackage{extsizes}
\usepackage[super,sort&compress,comma]{natbib} 
\usepackage[version=3]{mhchem}
\usepackage[left=1.5cm, right=1.5cm, top=1.785cm, bottom=2.0cm]{geometry}
\usepackage{balance}
\usepackage{mathptmx}
\usepackage{sectsty}
\usepackage{graphicx} 
\usepackage{lastpage}
\usepackage[format=plain,justification=justified,singlelinecheck=false,font={stretch=1.125,small,sf},labelfont=bf,labelsep=space]{caption}
\usepackage{float}
\usepackage{fancyhdr}
\usepackage{fnpos}
\usepackage[english]{babel}
\addto{\captionsenglish}{%
  
}
\usepackage{array}
\usepackage{droidsans}
\usepackage{charter}
\usepackage[T1]{fontenc}
\usepackage[usenames,dvipsnames]{xcolor}
\usepackage{setspace}
\usepackage[compact]{titlesec}
\usepackage{hyperref}
\usepackage{amsmath}
\usepackage{amssymb}
\usepackage{comment}

\definecolor{cream}{RGB}{222,217,201}

\usepackage{multirow}
\newcolumntype{C}{>{$}c<{$}} 

\begin{document}

\pagestyle{fancy}
\thispagestyle{plain}
\fancypagestyle{plain}{
\renewcommand{\headrulewidth}{0pt}
}

\makeFNbottom
\makeatletter
\renewcommand\LARGE{\@setfontsize\LARGE{15pt}{17}}
\renewcommand\Large{\@setfontsize\Large{12pt}{14}}
\renewcommand\large{\@setfontsize\large{10pt}{12}}
\renewcommand\footnotesize{\@setfontsize\footnotesize{7pt}{10}}
\makeatother

\renewcommand{\thefootnote}{\fnsymbol{footnote}}
\renewcommand\footnoterule{\vspace*{1pt}%
\color{cream}\hrule width 3.5in height 0.4pt \color{black}\vspace*{5pt}} 
\setcounter{secnumdepth}{5}

\makeatletter 
\renewcommand\@biblabel[1]{#1}            
\renewcommand\@makefntext[1]%
{\noindent\makebox[0pt][r]{\@thefnmark\,}#1}
\makeatother 
\renewcommand{\figurename}{\small{Fig.}~}
\sectionfont{\sffamily\Large}
\subsectionfont{\normalsize}
\subsubsectionfont{\bf}
\setstretch{1.125} 
\setlength{\skip\footins}{0.8cm}
\setlength{\footnotesep}{0.25cm}
\setlength{\jot}{10pt}
\titlespacing*{\section}{0pt}{4pt}{4pt}
\titlespacing*{\subsection}{0pt}{15pt}{1pt}

\fancyfoot{}
\fancyfoot[LO,RE]{\vspace{-7.1pt}\includegraphics[height=9pt]{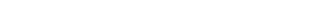}}
\fancyfoot[CO]{\vspace{-7.1pt}\hspace{11.9cm}\includegraphics{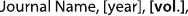}}
\fancyfoot[CE]{\vspace{-7.2pt}\hspace{-13.2cm}\includegraphics{head_foot/RF}}
\fancyfoot[RO]{\footnotesize{\sffamily{1--\pageref{LastPage} ~\textbar  \hspace{2pt}\thepage}}}
\fancyfoot[LE]{\footnotesize{\sffamily{\thepage~\textbar\hspace{4.65cm} 1--\pageref{LastPage}}}}
\fancyhead{}
\renewcommand{\headrulewidth}{0pt} 
\renewcommand{\footrulewidth}{0pt}
\setlength{\arrayrulewidth}{1pt}
\setlength{\columnsep}{6.5mm}
\setlength\bibsep{1pt}

\makeatletter 
\newlength{\figrulesep} 
\setlength{\figrulesep}{0.5\textfloatsep} 

\newcommand{\topfigrule}{\vspace*{-1pt}%
\noindent{\color{cream}\rule[-\figrulesep]{\columnwidth}{1.5pt}} }

\newcommand{\botfigrule}{\vspace*{-2pt}%
\noindent{\color{cream}\rule[\figrulesep]{\columnwidth}{1.5pt}} }

\newcommand{\dblfigrule}{\vspace*{-1pt}%
\noindent{\color{cream}\rule[-\figrulesep]{\textwidth}{1.5pt}} }

\makeatother

\twocolumn[
  \begin{@twocolumnfalse}
{
}\par
\vspace{1em}
\sffamily
\begin{tabular}{m{4.5cm} p{13.5cm} }

\includegraphics{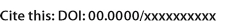} & \noindent\LARGE{\textbf{Sensitivity of solid phase stability to the interparticle potential range: studies of a new Lennard-Jones like model}} \\
\vspace{0.3cm} & \vspace{0.3cm} \\

 & \noindent\large{Olivia S. Moro,$^{\ast}$\textit{$^{ab}$} Vincent Ballenegger,\textit{$^{a}$} Tom L. Underwood\textit{$^{c}$} and Nigel B. Wilding\textit{$^{b}$$^{\ddag}$}} \\

\includegraphics{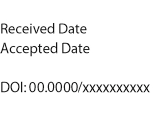} & \noindent\normalsize{
In a recent article, Wang {\em et al} (Phys. Chem. Chem. Phys., 2020, {\bf 22}, 10624) introduced a new class of interparticle potential for  molecular simulations. 
The potential is defined by a single range parameter, eliminating the need to decide how to truncate truly long-range interactions like the Lennard-Jones (LJ) potential.
 The authors explored the phase diagram for a particular value of the range parameter for which their potential is similar in shape to the LJ $12$-$6$ potential. We have reevaluated the solid phase behaviour of this model using both Lattice Switch Monte Carlo and thermodynamic integration. In addition to finding that the boundary between hexagonal close packed (hcp) and face centred cubic (fcc) phases presented by Wang {\em et al} was calculated incorrectly, we show that owing to its finite range, the new potential exhibits several `artifact' reentrant transitions between hcp and fcc phases. The artifact phases, which do not occur in the full (untruncated) LJ system, are also found for typically adopted forms of the truncated and shifted LJ potential. However, whilst in the latter case one can systematically investigate and correct for the effects of the finite range on the calculated phase behaviour, this is not possible for the new potential because the choice of range parameter affects the entire potential shape. Our results highlight that potentials with finite range may fail to represent the crystalline phase behavior of systems with long-range dispersion interactions, even qualitatively.

} 

\end{tabular}

 \end{@twocolumnfalse} \vspace{0.6cm}

  ]

\renewcommand*\rmdefault{bch}\normalfont\upshape
\rmfamily
\section*{}
\vspace{-1cm}


\footnotetext{\textit{$^{a}$~ Université de Franche-Comté, CNRS, Institut UTINAM, F-25000 Besançon, France.}}
\footnotetext{\textit{$^{b}$~HH Wills Physics Laboratory, Royal Fort, University of Bristol, Bristol BS8 1TL, U.K. }}
\footnotetext{\textit{$^{c}$~{Scientific Computing Department STFC, Rutherford Appleton Laboratory, Harwell Campus, Didcot, OX11 0QX, U.K.} }}
\footnotetext{\textit{$^{\ast}$~ Email: hn22404@bristol.ac.uk}}
\footnotetext{\textit{$^{\ddag}$~ Email: nigel.wilding@bristol.ac.uk}}




\section{Introduction}

\begin{figure}[h]
\centering
  \includegraphics[scale=0.55]{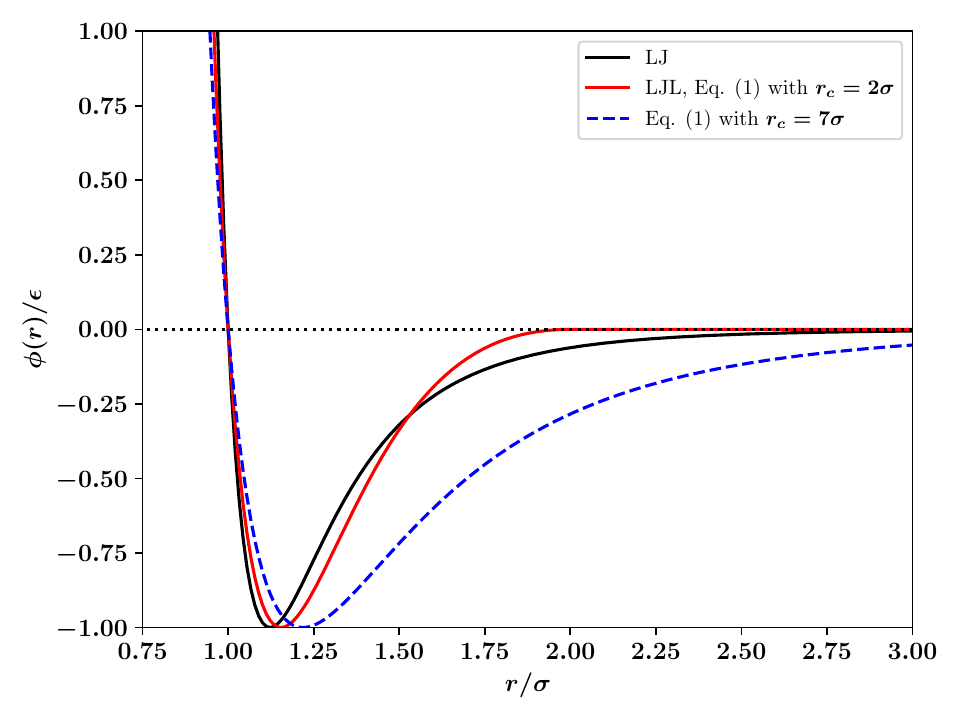}
  \caption{Comparison of the standard $12-6$ Lennard-Jones (LJ) interaction potential with the potential of Wang et al \cite{wang2020} (Eq.~\ref{eq:LJL}) with range parameters $r_c$ listed in the legend. The case $r_c=2.0\sigma$ is the Lennard-Jones like (LJL) potential.}
  \label{fgr:LJL_pot}
\end{figure}

In a recent paper, Wang {\em et al} \cite{wang2020} have proposed a new class of interparticle potential for use in molecular simulation. The new potential, which has received considerable attention, takes the form
\begin{equation}
\label{eq:LJL}
	\phi(r) \equiv 
    \begin{cases} \epsilon \alpha(r_{c}) \left[\left(\frac{\sigma}{r}\right)^{2}-1\right]\left[\left(\frac{r_c}{r}\right)^{2}-1\right]^{2}  & \text{for $r \leq r_c$,} \\
	  0  & \text{for $r > r_c$.}
	\end{cases}
\end{equation}
Here $\sigma$ sets the length scale, and
\begin{equation}
    \label{eq:alpha}
    \alpha(r_{c}) = \frac{27}{4}\left( \dfrac{r_{c}}{\sigma} \right)^{2}
    \left( \left( \dfrac{r_{c}}{\sigma}\right)^{2} -1 \right)^{-3}
\end{equation}
is a coefficient that ensures that the depth of the attractive well is $-\epsilon$.
The potential is formulated such as to vanish smoothly at the specified value of the cutoff parameter $r_c$, thus circumventing the question of which truncation scheme to employ when  seeking to render computationally tractable a truly long-ranged interaction such as the well known Lennard-Jones (LJ) potential. However, in contrast to a truncated LJ potential, the choice of $r_c$ in \eqref{eq:LJL} sets not just the truncation distance, but also determines the overall shape of the potential. The authors find that for $r_c=2\sigma$ (for which $\alpha=1$), its form is similar to that of the LJ potential as shown in Fig.~\ref{fgr:LJL_pot}. We shall henceforth refer to Eq.~(\ref{eq:LJL}) with $r_c=2\sigma$ as the Lennard-Jones like (LJL) potential.

Wang {\em et al} presented various thermophysical properties of the LJL potential including the phase diagram for the vapor, liquid and crystalline solid phases, all computed from free energy measurements via thermodynamic integration (TI). In the present work we have revisited the crystalline solid region of the LJL model using both TI and Lattice Switch Monte Carlo (LSMC). The latter is a powerful method for determining the free energy between coexisting crystalline phases. Our results reveal an error in the measurement of the phase boundary between hcp and fcc phase presented by Wang {\em et al} \cite{wang2020}. Moreover we find that on increasing number density the correct phase diagram for the solid region manifests several reentrant fcc and hcp phases --features that are absent in the full LJ potential. We trace the source of these re-entrant phases to the finite range of the LJL potential ie. the lack of long ranged dispersion interactions. Our results for the phase behaviour are set out in Sec.~\ref{sec:pd} while Sec.~\ref{sec:concs} provides a discussion and summarises our conclusions concerning the model.

\section{Phase diagram of the Lennard-Jones like model}
\label{sec:pd}
We have studied the crystalline solid phase behaviour of the LJL model using both Lattice Switch Monte Carlo and thermodynamic integration in conjunction with Molecular Dynamics simulation. We have also performed ground state energy calculations as a function of number density for both the LJL model and a truncated and shifted LJ potential. Implementation details for these methods and the results of our calculations are reported below and in Supplementary Material (SM) \cite{Moro2023SM}. In all cases, numerical values are expressed in dimensionless units, namely reduced temperature $T^{*}=k_{B}T/\epsilon$, pressure $p^{*}=p\sigma^{3}/\epsilon$ and particle number density $\rho^{*}=\rho\sigma^{3}$. For notational simplicity we shall henceforth suppress the superscript $^{*}$ on these quantities.

\subsection{Lattice Switch Monte Carlo calculations} 
\label{sec:2.1}

LSMC \cite{Bruce1997,Bruce2000,Jackson2002,Jacksonthesis} is a well established and powerful simulation technique for determining solid-solid coexistence parameters. Its high precision stems from the fact that it focuses on the difference in the free energy between two candidate stable phases rather than the absolute free energy of each~\cite{Bruce2003}.  In the course of a LSMC simulation the system switches repeatedly between two candidate stable structures allowing the accumulation of statistics on their relative statistical weight. To enable such switching, the sampling must be biased to visit -- on a regular basis -- certain 'gateway' configurations from which a switch from one phase to the other can be launched. The requisite bias function can be obtained by using the transition matrix method\cite{Smith1996,McNeil-Watson2006}. The effects of the bias are unfolded from the statistics {\em a posteriori}. 

We have deployed the LSMC method to obtain the crystalline solid phase behaviour of the LJL model using the implementation included within the general purpose MC simulation engine DL\_MONTE \cite{dlmonte}. Details regarding the simulations can be found in the SM \cite{Moro2023SM}. We investigate in particular the relative stability of  hcp and fcc structures as a function of the particle number density $\rho=N/V$ and temperature $T$. To do so, we work within the isobaric-isothermal (constant-NpT) ensemble for which LSMC provides direct access to the Gibbs free energy difference between the hcp and fcc phases $\Delta G = G_{hcp}-G_{fcc}$. This is given by

$$ \Delta G = -k_{B}T\log(P_{hcp}/P_{fcc})$$
where $P_{hcp}$ and $P_{fcc}$ are the integrated probabilities for the system to be found in microstates typical of hcp and fcc respectively.

Coexistence state points are defined by $\Delta G=0$. We located the value of the pressure ($p$) that corresponds to phase coexistence for a prescribed temperature using a root finding algorithm. Specifically, we applied Newton-Raphson's method to the function $\Delta G(p)$ which leads, since $dG = V dP$, to iterate the pressure according to\cite{Jacksonthesis}
\begin{equation}\label{eq:Pnew}
	p_{i+1} = p_{i} - \dfrac{ \Delta G_{i} }{ \Delta V_{i} },
\end{equation}
where $\Delta V_i$ is the volume difference between the two phases at the $i^{\rm th}$ iteration. At each iteration, we computed the LSMC bias function afresh, though it is possible to avoid doing so by deploying histogram reweighting techniques~\cite{McNeil-Watson2006}.

\begin{figure}[h]
\centering
  \includegraphics[scale=0.575]{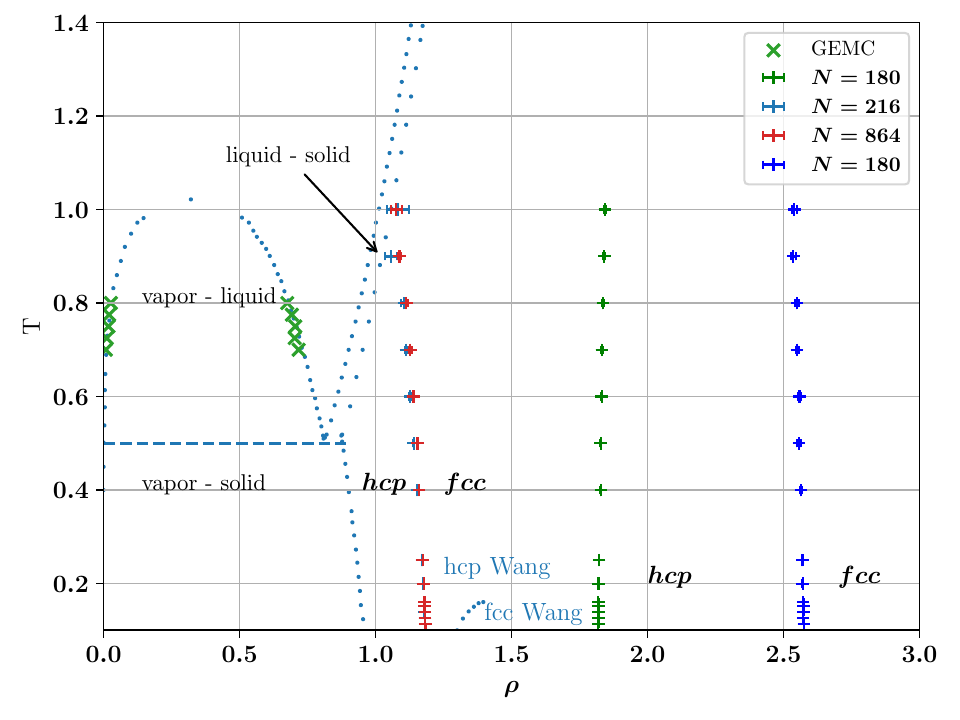}
  \caption{Phase diagram in the density-temperature plane of the LJL system, eq.~(\ref{eq:LJL}) with $r_c=2.0\sigma$. Blue dots are phase boundaries reported in Wang {\em et al}~\cite{wang2020}, symbols are the present work. Green crosses denote liquid-vapour coexistence densities calculated from GEMC. Our LSMC estimates of the reentrant hcp-fcc coexistence boundaries are denoted by $+$ and were obtained in the reduced temperature range $0<T\le 1$ and for the particle numbers $N$ shown in the legend. Note the complete disparity with the single hcp-fcc boundary calculated by Wang {\em et al},  as marked by blue dots in the solid region.  In all cases, the difference in the coexisting hcp and fcc densities is invisibly small on the scale of the graph. Uncertainties for the coexistence densities were calculated as described in the SM~\cite{Moro2023SM} and unless indicated by an error bar, do not exceed the symbol size.}
  \label{fgr:phase_diagram}
\end{figure}

Our LSMC results for the phase diagram are presented in Fig.~\ref{fgr:phase_diagram}. While we find agreement with the work of Wang {\em et al}\cite{wang2020} for the vapor-fluid binodal (which we have determined using separate Gibbs Ensemble Monte Carlo (GEMC) simulations within DL\_MONTE), our results for the crystalline region are very different. Specifically,  we find at least three separate lines of hcp-fcc transitions, none of which coincide with that of Wang {\em et al}. In order to check our findings and throw light on the discrepancy, we have performed TI calculations for the solid region. The results of these calculations are described in the next subsection.

\subsection{Thermodynamic integration calculations} 
\label{sec:2.2}
To elucidate the origin of the discrepancy between our prediction for the hcp-fcc coexistence line(s) based on LSMC calculations and the thermodynamic integration calculations of Ref.~\cite{wang2020}, we have computed absolute Helmholtz free energies of both phases using another simulation program, {\it GROMACS}\cite{GROMACS2005,gromacs,Moro2023SM}, and the TI method\cite{Frenkel1984} for a system comprising $N=768$ particles. We selected a number of ($T, \rho$) state points where absolute Helmoltz free energies have previously been calculated by Wang {\em et al.} and given in the supplementary material of Ref.~\cite{wang2020}.

For our TI calculations we adopted the integration path described by Aragones {\em et al}~\cite{aragones2012,aragones2013} to connect the crystalline phase of interest and a known reference state, namely the ideal Einstein crystal with one atom fixed at its lattice site (state 0)\footnote{Note that this is a different choice of integration path to that adopted by Wang {\em et al}\cite{wang2020}}.The Helmholtz free energy of state 0 is $A_0 = 3/2 (N-1) \ln(\beta k_E \Lambda_T^2/\pi)$ where $\Lambda_T=h/\sqrt{2\pi m k_B T}$ is the thermal de Broglie wavelength, $\beta=1/(k_BT)$ and $k_E$ is the spring constant of the $(N-1)$ harmonic oscillators. The absolute free energy $A$ can be expressed as~\cite{Vega2008,aragones2012}

\begin{equation}	    \label{AfromTI}
A = A_0 + \Delta A_1 + \Delta A_2 + k_B T \ln\Big(\frac{N \Lambda_T^3}{V}\Big)
\end{equation}
where $\Delta A_1$ is the free energy difference between state 1, which is an Einstein crystal where the spring-bound atoms interact through LJ-like interactions, and state 0. $\Delta A_2$ is the free energy difference between state 2, the crystalline phase of interest but with one atom fixed, and state 1. The last term in Eq.~\eqref{AfromTI} compensates for the fixed atom constraint. $\Delta A_1$ is computed using Bennett's formula $\Delta A_1 = - k_B T \ln \langle 
e^{-\beta (E_1-E_0)} \rangle_0$ where the thermal average is over equilibrium trajectories of the ideal Einstein crystal (state 0). Notice that $E_1 - E_0 = E_{\rm pair}$ is merely the sum of the LJ-like pair interactions. To avoid overflows in the exponent, the formula is rewritten as $\beta \Delta A_1 = \beta E_{\rm pair}^{(T=0)} - \ln \left\langle \exp(-\beta(E_{\rm pair} - E_{\rm pair}^{(T=0)})) \right \rangle_0$ where $E_{\rm pair}^{\rm (T=0)}$ is the sum of the pair interactions in the perfect crystal at $T=0$. The spring constants were adjusted so that the second term (per particle) $-\ln\langle \exp(-\beta(E_{\rm pair} - E_{\rm pair}^{(T=0)})) \rangle_0/N$  is close to $0.02$, as recommended~\cite{aragones2012,aragones2013}. The integral in the contribution $\Delta A_2 = -\int_0^{k_E}\Big\langle \sum_{i=1}^N \frac{1}{2}\Big(\vec{r}_i - \vec{r}_i^{(0)}\Big)^2\Big \rangle_{k_E'} dk_E',$ where $r_i^{(0)}$ is the reference lattice site of the $i^{\rm th}$ particle and the thermal average is over an interacting Einstein crystal with spring constant $k_E'$, was computed by using the change of variable~\cite{Frenkel1984,Vega2008} $x = \ln(\beta\frac{1}{2}  k'_E (10^{-1}\,{\rm nm})^2 + e^{3.5})$ and a Gauss-Legendre quadrature with $15$ points.

The results of our TI calculations are compared in Table~\ref{tbl:deltaA} with LSMC calculations that we have performed in the canonical (constant-$NVT$) ensemble. The state points listed are a selection of those also studied by Wang {\em et al}~\cite{wang2020} (see their SM). At $T=0.1$, our LSMC and TI both predict a (first) transition from hcp to fcc around density $\rho \approx 1.18(1)$ (where $\Delta A(\rho)$ changes sign), far removed from the density $\rho \approx 1.29$ at which coexistence is predicted in Ref.~\cite{wang2020}.  At higher temperatures we find consistency between our canonical LSMC and TI estimates of transition densities on this first hcp-fcc phase boundary, and the constant-NpT LSMC calculations of the full phase boundary reported in Fig.~\ref{fgr:phase_diagram}.  While error bars are difficult to assess in the TI method, they can be calculated straightforwardly\cite{Moro2023SM} for LSMC and standard deviations on the LSMC results are reported in the table.  Absolute free energies depend on the atomic mass via the thermal de Broglie wavelength $\Lambda_T$, while the free energy difference $\Delta A$ does not. Our TI values were computed by setting $\Lambda_T = \sigma$, separately at each considered temperature. This standard convention was apparently also used in Ref.~\cite{wang2020}.

At all densities and temperatures reported in Table~\ref{tbl:deltaA}, the free energy difference $\Delta A = A_{\rm hcp} - A_{\rm fcc}$ that we obtain with TI deviates slightly and systematically from our LSMC results by a shift of the order $-0.02 N k_B T$. This small temperature-independent shift might be due to systematic integration errors in the TI calculations. Much larger discrepancies are observed between our LSMC and TI values for $\Delta A$ and those of Ref.~\cite{wang2020} (compare columns $3$ and $4$ with column $5$ of Table~\ref{tbl:deltaA} respectively). Comparing separately the free energies estimates of the fcc and hcp phases, we find generally reasonable agreement between our TI results and those of Wang {\em et al}\cite{wang2020} for the fcc phase, but a much larger discrepancy for the hcp phase, as reported in the rightmost two columns of Table~\ref{tbl:deltaA}. This shows that the absolute free energies of the hcp phase at least have very likely been miscalculated in Ref.~\cite{wang2020}, leading to a very different and incorrect crystalline phase boundary. Since the missing contribution is proportional to the temperature, the hcp-fcc coexistence curve determined in Ref.~\cite{wang2020} (for $\rho \leq 1.4$) deviates increasingly from our coexistence curve.  We note that the error in the calculation of the free energy also raises questions regarding the correctness of the estimates of the liquid-solid phase boundary reported by Wang {\em et al}, as shown in Fig.~\ref{fgr:phase_diagram}, though we have not checked that particular calculation. 

\begin{table*}[h]
\centering
\small
\caption{Free energy difference $\Delta A$, in units of $Nk_{\rm B}T$, between the hcp and fcc structures of the LJL solid ($r_c = 2\sigma$) at a selection of state points. The discrepancy between the LSMC and TI calculations is small (see 6$^{th}$ column) and might be due to systematic integration errors in the TI method. The difference between the absolute free energy calculated via TI in this work and that in Ref.~\cite{wang2020} are shown for both the fcc and the hcp structures in the rightmost two columns. Note the large discrepancy for the hcp phase.}
\label{tbl:deltaA}
  \begin{tabular*}{0.88\textwidth}{@{\extracolsep{\fill}}*8C}
    \hline
    \multirow{2}{*}{$T$} & \multirow{2}{*}{$\rho$}  & \multicolumn{3}{c}{ $\large \Delta A = A_{\rm hcp} - A_{\rm fcc}$}
    & \text{Difference}
    & \multicolumn{2}{c}{ $A\text{\footnotesize(this work)} - A\text{\footnotesize(Ref.~\cite{wang2020})}$} 
    \\
      &         &  {\footnotesize \text{LSMC}} & {\footnotesize \text{TI}} & {\footnotesize \text{Ref.~\cite{wang2020}} } 
      & \text{\footnotesize LSMC} -  \text{\footnotesize TI}
      & {\footnotesize \text{fcc}} & {\footnotesize \text{hcp}} 
      \\ 
      \hline
0.1\phantom{00} & 1.16 & -0.044(4)\phantom{000} & -0.021 & -0.228 & -0.02  & 0.02 & 0.23\\ 
0.1\phantom{00} & 1.20 & 0.027(3)\phantom{0} & 0.054 & -0.157 & -0.03 & 0.02 & 0.23  \\ 
0.1\phantom{00} & 1.24 & 0.106(10) & 0.131 & -0.083 & -0.03 & 0.02 & 0.23  \\ 
0.1\phantom{00} & 1.28 & 0.181(17) & 0.206 & -0.012 & -0.02 & 0.02 & 0.23  \\ 
0.1\phantom{00} & 1.32 & 0.240(23) & 0.264 & 0.020 & -0.02 & 0.01 & 0.23  \\ 
0.204 & 1.2\phantom{0} & 0.0185(17) & 0.036 & -0.168 & -0.02 & 0.02 & 0.22 \\ 
0.308 & 1.2\phantom{0} & 0.0147(14) & 0.032 & -0.172 & -0.02 & 0.00 & 0.20 \\ 
0.399 & 1.2\phantom{0} & 0.0135(13) & 0.030 & -0.175 & -0.02 & 0.01 & 0.22 \\ 
0.503 & 1.2\phantom{0} & 0.0120(13) & 0.026 & -0.178 & -0.01 & -0.01 & 0.19 \\ 
0.607 & 1.0\phantom{0} & -0.0107(11) & -0.0054\phantom{0}   &    -0.206    &  -0.01     &
  -0.07    &   0.13   \\
0.607 & 1.1\phantom{0} & -0.0053(6)\phantom{00} &   0.0024    &    -0.204    &   -0.01    &  -0.04     &    0.17  \\
0.607 & 1.2\phantom{0} & 0.0109(11) & 0.028 & -0.181 & -0.02 & -0.02 & 0.19 \\
0.802 & 1.0\phantom{0} & -0.0046(3)\phantom{00} &  -0.0060\phantom{0}    &   -0.207    &  0.001   &    -0.05  &  0.15  \\
0.802 & 1.1\phantom{0} & -0.0007(1) &  0.0076  & -0.206   & -0.01     & -0.05  & 0.16\\
0.802 & 1.2\phantom{0} & 0.0102(10) & 0.018 & -0.187 &  -0.01 & -0.03  & 0.18\\
    \hline
  \end{tabular*}
\end{table*}

\subsection{Ground state energy calculations.}
\label{sec:2.3}

 The corrected $\rho-T$ phase diagram for the LJL model displayed in Fig.~ \ref{fgr:phase_diagram} shows that the boundaries between hcp and fcc phases are almost linear and independent of temperature over a wide temperature range which extends down to very low $T$. It is thus reasonable to assume that coexistence properties calculated from ground state (ie. $T=0$) energy calculations may aid an understanding of the overall solid state phase behaviour. 
 
 In the ground state, particles occupy their lattice sites, there are no entropic contributions to the hcp-fcc free energy difference, and energy calculations suffice to determine the phase behaviour.  We have used the potential eq.~(\ref{eq:LJL}) to calculate the ground state energy difference per particle $\Delta E/N = (E_{hcp}-E_{fcc})/N$ as a function of $\rho$. The criterion $\rho(\Delta E=0)$ corresponds to hcp-fcc transition points. Note though that while the stability of each phase is correctly predicted by the sign of $\Delta E$, the hcp and fcc phases have slightly different specific volumes. The coexisting densities are then not given merely by the condition $\Delta E = 0$, but by $\Delta G = \Delta E + p \Delta V = 0$. They can be deduced from the curve $E(\rho,T=0)$ of each phase by performing the common tangent construction, which imposes a common pressure at coexistence.  However as the coexistence densities straddle the density $\rho(\Delta E=0)$ and are very close in value to one another --indistinguishably so on the scale of fig.~\ref{fgr:phase_diagram}-- it is sufficient in the current context to simply determine $\rho(\Delta E=0)$.
 
\begin{figure}[h]
 \centering
 \includegraphics[scale=0.59]{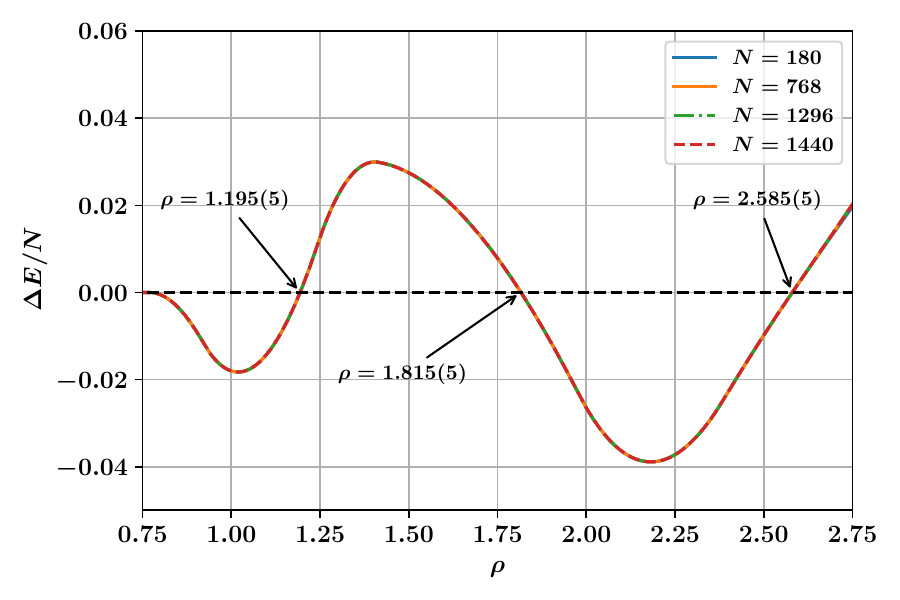}
 \caption{Dependence on number density $\rho$ of the ground state energy difference per particle $\Delta E/N = (E_{hcp}-E_{fcc})/N$  between hcp and fcc crystalline phase for the LJL potential. Data is shown for a range of particle number~$N$.  The results illustrate the reentrant hcp phase at $\rho=1.815(5)$, while the accord between the curves for different particle number $N$ demonstrates that the behaviour is not a finite-size effect.}
 \label{fgr:dE_T0K}
\end{figure}

 Our ground state energy calculations for the LJL model are shown in figure \ref{fgr:dE_T0K}, which plots $\Delta E/N$ as a function of $\rho$. One sees that $\Delta E = 0$ occurs for $\rho= 1.195(5)$ ($p=12.48$), which is close to the extrapolation to $T=0$ of the first hcp-fcc transition line shown in fig.~\ref{fgr:phase_diagram}. The same is true of the second and third transition lines shown in fig.~\ref{fgr:phase_diagram} which our $T=0$ calculations place at densities of $\rho \approx 1.815(5)$ ($p=145.31$) and $\rho \approx 2.585(5)$ ($p=623.81$) respectively. Thus our ground state energy calculations are consistent with the phase boundaries found at finite temperatures via LSMC. 

\begin{figure}[h]
 \centering
 \includegraphics[scale=0.59]{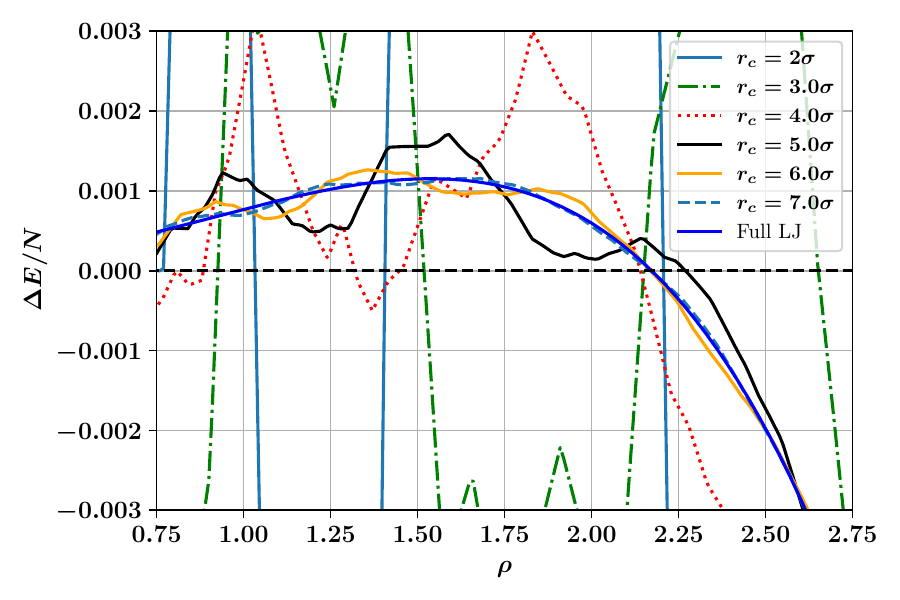}
 \caption{Dependence on number density $\rho$ of the ground state energy difference per particle between hcp and fcc crystalline phases as calculated for the truncated and shifted LJ potential at a selection of cutoff values. Also shown for comparison is the limiting case for the full (untruncated) LJ potential \cite{Stillinger2000}. Note how the results for the finite cutoff approach the limiting case for $r_c\gtrsim 7.0\sigma$.}
 \label{fgr:dE_T0K_LJts}
\end{figure}

It is instructive to compare the ground state phase behaviour of the LJL model with that of the truncated and shifted LJ potential which has the form

\begin{equation}
\label{eq:LJts}
	\tilde{\phi}_{LJ}(r) \equiv 
    \begin{cases} \phi_{LJ}(r) -\phi_{LJ}(r_c) & \text{for $r \leq r_c$} \\
	  0  & \text{for $r > r_c$},
	\end{cases}
\end{equation}
where $\phi_{LJ}(r)=4\epsilon  \left[\left(\frac{\sigma}{r}\right)^{12}-\left(\frac{\sigma}{r}\right)^{6}\right]$ is the full LJ potential, a portion of which is shown in fig.~\ref{fgr:LJL_pot}. We have used eq.~(\ref{eq:LJts}) to calculate the ground state energy as a function of the density for various values of the LJ truncation distance $r_c$. The results, which replicate and extend to higher values of $r_c$ similar calculations by Jackson {\em et al}\cite{Jackson2002}, are shown in figure \ref{fgr:dE_T0K_LJts}. This figure also includes the results of calculations for the limiting case of the full LJ potential given by  Stillinger~\cite{Stillinger2000} for which it is important to note there is only a {\em single} hcp-fcc transition, which occurs at $\rho=2.1728$.  On increasing $r_{c}$, our results for $\Delta E/N$ approach the limiting curve, but only for a large truncation distance of $r_c\gtrsim 6\sigma$. For smaller cutoffs $r_c\lesssim 5\sigma$, $\Delta E$ fluctuates wildly and changes sign at several densities, indicating that in this regime additional hcp-fcc transitions arises which are artifacts in the sense that they are not characteristic of the full LJ potential.

\begin{figure}[h]
 \centering
 \includegraphics[scale=0.59]{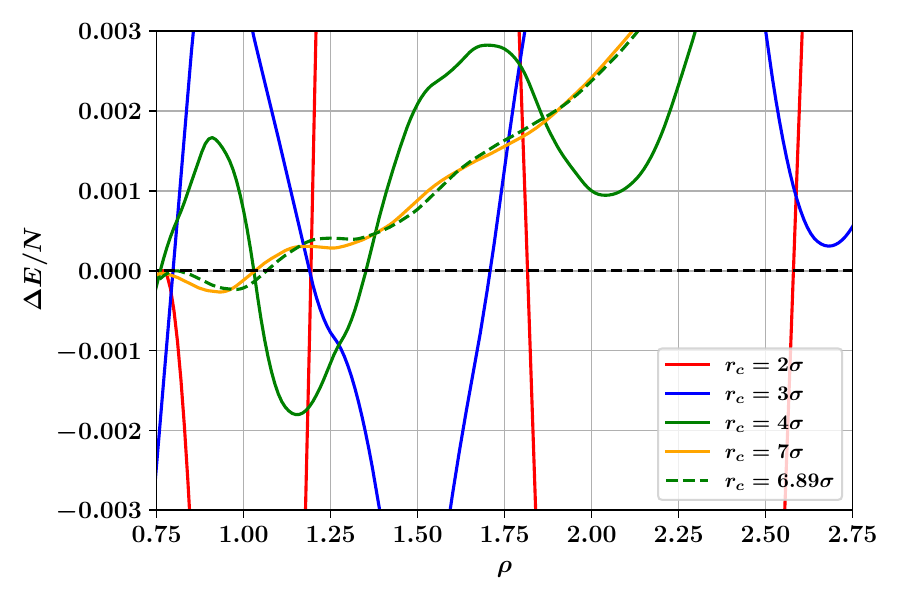}
 \caption{Dependence on number density $\rho$ of the ground state energy difference per particle between hcp and fcc crystalline solid phases as calculated for the LJL potential of Eq.~\ref{eq:LJL} at a selection of values of the range parameter $r_{c}$. For solid like densities ($\rho>0.8$), the smallest value of the range parameter for which only a single hcp-fcc transition occurs is $r_c=6.89\sigma$.}
 \label{fgr:dE_T0K_LJL_rc}
\end{figure}

 We have also considered the changes to the $T=0$ phase behaviour of the potential of Wang {\em et al} that result from varying the range parameter $r_c$ in Eq.~\ref{eq:LJL}. The results displayed in fig.~\ref{fgr:dE_T0K_LJL_rc} show that the number and location of phases occurring varies dramatically, similarly to what was found in fig.~\ref{fgr:dE_T0K_LJts} for the truncated and shifted LJ potential. In order for a single hcp-fcc transition to occur, as in the full LJ potential, one must utilise in Eq.~\ref{eq:LJL} a range parameter $r_c\gtrsim 7\sigma$. However, as  shown in Fig.~\ref{fgr:LJL_pot} and discussed below, for this value of the range parameter, the potential of Eq.~\ref{eq:LJL} is not at all Lennard-Jones like.

\section{Discussion and conclusions}
\label{sec:concs}

In this work we have reconsidered the phase diagram of a recently proposed finite-ranged potential, Eq.~(\ref{eq:LJL}), that has been recommended as an alternative to the use of truncated LJ potentials in molecular simulation \cite{wang2020}. The advantage of this LJL potential is that while it is similar in shape to the LJ potential, it is defined in such a way as to vanish smoothly at the prescribed value of the range parameter. It thus provides a unique truncation scheme, which contrasts with the multiple approaches in common use for the LJ potential. 

We have studied the phase diagram of the LJL model both in the liquid-vapor and the crystalline solid regions. While we confirm the original authors' findings for the vapor-liquid binodal, our results in the crystalline phases deviate greatly. Specifically on increasing the number density starting from the solid branch of the vapor-solid coexistence region, we find a very different hcp-fcc boundary to that reported. Our calculated hcp-fcc boundary (fig.~\ref{fgr:phase_diagram}) exhibits relatively weak temperature dependence and intersects the freezing line at a fluid-hcp-fcc triple point. We have traced the discrepancy between our results and those of Wang {\em et al} to an apparently incorrect calculation of the free energy on their part.

 We find, moreover, that on increasing the density further within the crystalline region, the LJL potential exhibits not one but multiple reentrant hcp-fcc transitions that extend over a wide range of temperatures down to $T=0$. Indeed a previous study by Choi {\em et al}\cite{Choi1991} of the LJ potential, truncated and shifted at $r_c=2.5\sigma$, reported a qualitatively similar phase diagram, displaying reentrant hcp and fcc transitions whose boundaries depend only weakly on temperature, as well as a fluid-hcp-fcc triple point. Importantly, however, such features are {\em absent} in the full LJ potential which except for a very small temperature range exhibits only a single hcp-fcc transition line, which is strongly temperature dependent at low $T$ and does not intersect the melting curve~\cite{Jackson2002,Jacksonthesis,Travesset2014}. Thus the solid phase diagrams of both the LJL potential and the truncated and shifted LJ system exhibit phases that are artifacts in the sense that they do not occur in the full LJ potential.

It is well established that the difference in free energy between hcp and fcc phases for spherically symmetric interparticle potentials can be small across a range of state points~\cite{Frenkel1984,Bruce2000,Jackson2002}. Previous studies of the ground state ($T=0$) energy as a function of number density of the truncated and shifted LJ potential have shown that artifact transitions arise due to the sensitivity of the sign of the hcp-fcc energy difference $\Delta E/N$ to the truncation range~\cite{Jackson2002,Jacksonthesis}. We note also that the magnitude of $\Delta E/N$ for a truncated potential can be much larger than for the full LJ potential. For instance, when truncating the LJ interaction at some short cutoff, say $r_c = 2\sigma$, $\Delta E/N$ is about $10$ times that of the full LJ potential. The size of this energy difference would appear to be responsible for the insensitivity to temperature of the measured hcp-fcc coexistence boundaries, as seen for both the truncated LJ potential~\cite{Choi1991} and the LJL system (fig.~\ref{fgr:phase_diagram}), because it dominates over the hcp-fcc entropy difference. It is therefore not surprising that the much smaller value of $\Delta E$ associated with the full LJ potential leads to very different crystalline phase diagram \cite{Jackson2002,Travesset2014}. Ground state energy calculations for the truncated and shifted LJ potential performed in the present work (fig.~\ref{fgr:dE_T0K_LJts}) demonstrate that, at least at $T=0$, one requires $r_c\gtrsim 6\sigma$ in order to eliminate artifact phases. This would therefore seem to represent a lower bound on the truncation length scale necessary to obtain a solid phase diagram in qualitative agreement with that of the full LJ potential\footnote{Note, however, the finding of Jackson {\em et al} ~\cite{Jackson2002,Jacksonthesis} that by treating the ground state exactly (ie in untruncated form) so that the effects of the potential truncation are confined to the fluctuation spectrum, one can obtain a finite-temperature phase diagram close to that of the full LJ potential for shorter truncation distances.}. 

 In the case of the LJL potential of Wang {\em et al}, Eq.~\ref{eq:LJL}, we find (see fig.~\ref{fgr:dE_T0K_LJL_rc}) that the number of stable solid phases occurring at $T=0$ can similarly be reduced by increasing the range parameter $r_c$, such that for $r_c\gtrsim 7\sigma$ only a single hcp-fcc transition remains.  However, as fig.~\ref{fgr:LJL_pot} shows, for such large values of the range parameter, the potential is no longer similar to the LJ potential. This latter point exposes a significant disadvantage of the LJL potential, Eq.~\eqref{eq:LJL}: whilst for a truncated LJ system one can investigate (and systematically correct for~\cite{Jacksonthesis,Jackson2002}) the effects of the finite potential range when modelling a given physical system, this seems impossible for the LJL potential because the range parameter controls not only the truncation distance but also the entire potential shape. Thus while the LJL potential may be perfectly adequate in situations where one wishes to employ a simple truncated attractive interparticle interaction, it comes with a caveat on its use for crystalline solid phases since it may be unable to represent in a qualitatively correct form the phase behaviour of real atomic systems whose overall features are determined in significant part by long-ranged dispersion interactions.

\section*{Author Contributions}

All authors were involved in conceiving and directing the research. OM and VB performed the numerical calculations. All authors wrote the paper. 

\section*{Conflicts of interest}
There are no conflicts to declare.

\section*{Acknowledgements}
NBW thanks Francesco Turci for drawing attention to a relevant reference and R. Evans for helpful comments on the manuscript. The computer simulations were carried out using the computational facilities of the Institute UTINAM and of the Mésocentre de calcul de Franche-Comté . This work was partly funded by UBFC within the project I-SITE BFC ``STABHYDRA''.

\appendix

\balance


\bibliography{rsc} 
\bibliographystyle{rsc} 

\end{document}